\documentclass[aps,prb,twocolumn,superscriptaddress]{revtex4-2}
\usepackage{graphicx}
\usepackage{graphics}
\usepackage{color}
\usepackage{version}

\usepackage{bm}
\usepackage{textcomp}
\usepackage{xfrac}

\DeclareMathAlphabet{\mathit}{OT1}{ptm}{m}{it}

\begin{document}

\excludeversion{details}


\title{Electronic correlations and spin frustration in the molecular conductors $\kappa$-(BEDT-TTF)$_2$X probed by magnetic quantum oscillations}

\author{S. Erkenov}
\affiliation{Walther-Mei{\ss}ner-Institut, Bayerische Akademie der Wissenschaften, D-85748 Garching, Germany}
\affiliation{School of Natural Sciences, Technische Universit{\"a}t M{\"u}nchen, D-$85748$ Garching, Germany}

\author{S. Fust}
\affiliation{Walther-Mei{\ss}ner-Institut, Bayerische Akademie der Wissenschaften, D-85748 Garching, Germany}
\affiliation{School of Natural Sciences, Technische Universit{\"a}t M{\"u}nchen, D-$85748$ Garching, Germany}

\author{S. Oberbauer}
\affiliation{Walther-Mei{\ss}ner-Institut, Bayerische Akademie der Wissenschaften, D-85748 Garching, Germany}
\affiliation{School of Natural Sciences, Technische Universit{\"a}t M{\"u}nchen, D-$85748$ Garching, Germany}

\author{W. Biberacher}
\affiliation{Walther-Mei{\ss}ner-Institut, Bayerische Akademie der Wissenschaften, D-85748 Garching, Germany}

\author{N. D. Kushch}
\affiliation{Institute of Problems of Chemical Physics, Russian Academy of Sciences, Chernogolovka, 142432 Russian Federation}

\author{H. M\"{u}ller}
\affiliation{ESRF - The European Synchrotron, F-38043 Grenoble 9, France}

\author{F. L. Pratt}
\affiliation{ISIS Neutron and Muon Source, STFC Rutherford Appleton Laboratory, Chilton, Didcot OX11 0QX, United Kingdom}

\author{R. Gross}
\affiliation{Walther-Mei{\ss}ner-Institut, Bayerische Akademie der Wissenschaften, D-85748 Garching, Germany}
\affiliation{School of Natural Sciences, Technische Universit{\"a}t M{\"u}nchen, D-$85748$ Garching, Germany}
\affiliation{Munich Center for Quantum Science and Technology (MCQST), D-80799 Munich, Germany}

\author{M. V. Kartsovnik}
\email{mark.kartsovnik@wmi.badw.de}
\affiliation{Walther-Mei{\ss}ner-Institut, Bayerische Akademie der Wissenschaften, D-85748 Garching, Germany}

\date{\today}

\begin{abstract}
The layered molecular conductors $\kappa$-(BEDT-TTF)$_2$X are a perfect experimental platform for studying the physics of the Mott transition and related exotic electronic states. In these materials, the subtle balance between various instabilities of the normal metallic state can be efficiently changed by applying a very moderate external pressure or by subtle chemical modifications, e.g. by a replacement of the insulating anion X$^{-}$, frequently referred to as ``chemical pressure''. A crucially important but still unsettled issue is an exact understanding of the influence of physical and chemical pressure on the electronic structure. Here, we use magnetic quantum oscillations to explore in a broad pressure range the behavior of the key parameters governing the Mott physics, the electronic correlation strength ratio $U/t$ and the spin frustration ratio $t'/t$ in two $\kappa$ salts, the ambient-pressure antiferromagnetic insulator with X = Cu[N(CN)$_2$]Cl and the ambient-pressure superconductor with X = Cu(NCS)$_2$.
Our analysis shows that pressure effectively changes not only the conduction bandwidth but also the degree of spin frustration, thus weakening both the electronic correlation strength and the magnetic ordering instability. At the same time, we find that the replacement of the anion Cu[N(CN)$_2$]Cl$^-$ by Cu(NCS)$_2^-$ results in a significant increase of the frustration parameter $t'/t$,  leaving the correlation strength essentially unchanged.
\end{abstract}

\maketitle
\section{Introduction}
Layered organic charge-transfer salts have been extensively employed as model systems for exploring the correlation-driven Mott insulating instability and a plethora of associated fascinating phenomena ranging from conventional \cite{kano97,kano04,powe06,seo04,mori16,oshi17,okaz13} and exotic \cite{ried21,lunk12,gati18,kaga13,sasa17} charge- and spin-ordered states, to quantum spin and electric-dipole liquids \cite{kano11,shim16,kano17,shim17,hass18,urai22}, to valence-bond glass or solid \cite{ried19,shim07,miks21,pust22} phases as well as unconventional superconductivity emerging in the direct proximity to an insulating ground state \cite{powe11,arda12,clay19,toyo07}.
Of particular interest is the family $\kappa$-(BEDT-TTF)$_2$X, where BEDT-TTF stands for the radical-cation bis(ethylenedithio)tetrathiafulvalene forming conducting layers alternating with insulating layers of a monovalent anion X$^{-}$ \cite{toyo07,ishi98}.
The organic molecules in the layers form an anisotropic triangular lattice of dimers with the on-site (intra-dimer) Coulomb repulsion significantly exceeding the nearest- and next-nearest-neighbor (inter-dimer) transfer integrals, $U \gg t,t'$, see refs. \cite{hott03,powe11,kano11,kand09,kore14} for a review and inset in Fig.\,\ref{tt}(a) below for illustration. This gives rise to a narrow, effectively half-filled conducting band.
Most of the abovementioned electronic states can be realized in these compounds, depending on subtle details of their crystal and electronic band structure, which can be controlled, e.g., by applying a moderate pressure, typically below $1$\,GPa, or by modifying the insulating anion.
Pressure is known to reduce the electronic correlation strength ratio $U/t$ through increasing the conduction bandwidth, without changing the band filling. Therefore, the pressure-induced transition between the metallic and insulating ground states is generally referred to as a bandwidth-controlled metal-insulator transition (MIT).

The anion replacement in the $\kappa$ salts has also been widely believed to primarily modify the bandwidth and therefore considered as ``chemical pressure'', see, e.g.,
 \cite{ishi98,mori99a,powe11,kano11,arda12,pust18}.
This interpretation has, however, been questioned by first-principles band structures calculations \cite{kore14}, which suggested that the overall ground state properties of these materials are controlled by the degree of anisotropy of the dimer triangular lattice rather than by the correlation strength ratio $U/t$. The anisotropy of the triangular lattice, quantified by the ratio $t'/t$, is one of the key parameters in the physics of the Mott transition. Being directly relevant to the spin frustration, it is crucially important for the magnetic properties of the Mott-insulating state and for the nature of the eventual superconducting state in the adjacent domain of the phase diagram \cite{kano11,powe11,zhou17,wiet21,misu17}. Through its impact on the magnetic ordering instability it should  also influence the critical electronic correlation strength required for the MIT, see, e.g., \cite{wata06,ohas08,misu17}.

The first experimental argument in support of the theoretical prediction \cite{kore14} was obtained in the recent comparative study of magnetic quantum oscillations in two $\kappa$-(BEDT-TTF)$_2$X salts with X = Cu[N(CN)$_2$]Cl and Cu(NCS)$_2$ (hereafter referred to as $\kappa$-Cl and $\kappa$-NCS, respectively) under pressure \cite{ober23}. These salts have very similar electronic band structures, but different ambient-pressure ground states \cite{toyo07,ishi98}. $\kappa$-Cl is an archetypal antiferromagnetic Mott insulator, which transforms into a metal under a pressure of $20-40$\,MPa.  By contrast, the $\kappa$-NCS salt is already metallic at ambient pressure. Should the correlation strength be different in these salts, it must be reflected in the many-body renormalization of the effective mass \cite{brin70,geor96}.
However, the recent experiment \cite{ober23} has revealed no difference between the effective masses of the two salts in the pressure interval
$40 \lesssim p \lesssim 100$\,MPa, corresponding to the close proximity to the MIT in $\kappa$-Cl, on the metallic side of its phase diagram. This result suggests that the mass renormalization, hence the correlation strength ratio $U/t$ is approximately the same for both salts.
Given the virtually equal electronic correlation strength, the difference in the ambient-pressure ground states is natural to attribute to a difference in the spin frustration ratio $t'/t$. This would be fully consistent with the band structure calculations predicting a stronger frustration for the more metallic, though not weaker correlated, $\kappa$-NCS salt \cite{kore14,kand09}. However, for a conclusive proof, it is important to provide an experimental test for the $t'/t$ ratio in the two salts. Further, for a more accurate comparison of the many-body renormalization effects, an experimental data on the effective masses in a considerably broader pressure range is needed.

To this end, we have carried out a detailed study of magnetic quantum oscillations of interlayer magnetoresistance of the $\kappa$-Cl and $\kappa$-NCS salts in a broad range of pressures, up to $p\approx 1.5$\,GPa. This range covers both the close neighborhood of the MIT in $\kappa$-Cl and the region deep in the normal metallic state, where the electronic correlations are significantly reduced. The data obtained allow us to evaluate both the electronic correlation strength and the degree of spin frustration and to trace their evolution with pressure. In this way, we have obtained a quantitative information on the influence of pressure and anion substitution on the Mott-insulating and magnetic-ordering instabilities in the two prominent members of the $\kappa$-(BEDT-TTF)$_2$X family.

The paper is organized as follows. The next section describes the experimental details and conditions. The experimental results and their discussion are given in Sec.\,\ref{Res}. We start with the general behavior of the quantum oscillations of magnetoresistance (Shubnikov-de Haas, SdH oscillations) in the $\kappa$-Cl salt and its evolution with pressure.
In particular, we present here some details on the oscillation amplitude, possibly related to a pressure-dependent spin-splitting, and on the influence of the weak Fermi surface warping in the interlayer direction. For the $\kappa$-NCS salt, there is a vast literature on its high magnetic field properties, including quantum oscillations, see, e.g., refs. \cite{wosn96,toyo07,sing00,kart04,audo16} for a review.
Therefore for this salt we only give a very brief account of the SdH oscillations in the Supplemental Material \cite{sm-hp}, illustrating how the cyclotron masses were evaluated.
In Sect. \ref{freq} we present detailed data on the pressure dependence of the SdH frequencies and analyze them in terms of the effective dimer model \cite{caul94,prat10b}. We show that both the anion replacement and the pressure lead to significant changes in the inplane anisotropy reflected, in particular, in the spin frustration ratio $t'/t$. In Sec.\,\ref{mass}, the cyclotron effective masses corresponding to the two fundamental SdH frequencies are presented for both salts and compared with each other. Throughout the entire pressure range studied, both salts show very similar mass values. Moreover, the overall mass behavior is remarkably well described by the Brinkman-Rice model, indicating the electron-electron interactions as a dominant mechanism of the pressure-dependent mass renormalization and allowing its quantitative analysis.
Our conclusions are summarized in Sec.\,\ref{conc}.

\section{Experimental}
The single crystals used in the experiment were grown electrochemically. The $\kappa$-Cl crystals were grown following the procedure described in literature \cite{ishi98,will90,uray88}, using the carefully purified commercial BEDT-TTF donor. For the $\kappa$-NCS crystals, the BEDT-TTF donor was synthesized according to the specially developed protocol \cite{muel93,muel97,muel22} using [1,3]-dithiolo-[4,5-d][1,3-dithiole]-2,5-dione (thiapendione, TPD) as a starting compound. This procedure yielded crystals of very high quality (see the the Supplemental Material \cite{sm-hp}), which was particularly important for the observation and quantitative analysis of the high-frequency ($F_{\beta}$) SdH oscillations.

For the measurements, two
crystals of each salt, $\kappa$-Cl and $\kappa$-NCS, have been selected.
SdH oscillations were studied in the interlayer transport geometry, that is conventional for the layered organics \cite{kart04,sing00}. The resistance across the layers was measured using the standard four-probe low-frequency ($f \sim 10 - 300$\,Hz) a.c. technique. Annealed $20\,\mu$m-thick platinum wires, serving as current and voltage leads, were attached to the samples with a conducting graphite paste yielding the contact resistance $\sim 10-30\,\Omega$ at low temperatures.
The samples were mounted in a Be-Cu clamp pressure cell and cooled down in $^3$He or $^4$He variable-temperature inserts placed into a 15\,T superconducting solenoid.
The pressure $p$ was evaluated at room temperature and at 15\,K using a calibrated $n$-doped InSb pressure gauge (see the Supplemental Material to ref. \cite{ober23} for details). In the high-pressure range, above 0.8\,GPa, the calibration was crosschecked using the $p$-linear resistance of a manganin wire \cite{nomu79} as a reference. In what follows, all the indicated pressure values are those determined at $T =15$\,K. The error in the pressure determination did not exceed $10$\,MPa at $p < 0.2$\,GPa and $ 5\%$ at higher pressures.

All the measurements on $\kappa$-Cl sample \#1 and on both $\kappa$-NCS samples were done in a magnetic field applied perpendicular to the plane of conducting layers (crystallographic $ac$-plane and $bc$-plane for $\kappa$-Cl and $\kappa$-NCS, respectively). This is a conventional geometry for probing the inplane charge dynamics in layered materials as the field induces cyclotron orbits in the layer plane. The $\kappa$-Cl sample \#2 was used partly in the same geometry at lower pressures, $p < 0.4$\,GPa, whereas for the rest of the measurements this sample was tilted by an angle of $\theta = 25^{\circ}$ from the perpendicular field direction, as explained in Sec.\,\ref{Gen}. Taking into account the quasi-2D character of the present materials, we simply multiply the SdH frequencies and cyclotron masses, determined in the tilted fields, by the factor $\cos \theta$ to obtain the values corresponding to the perpendicular field.

\section{Results and discussion} \label{Res}
\subsection{SdH oscillations in $\kappa$-Cl: general features} \label{Gen}

\begin{figure}[tb]
\center
\includegraphics[width = 0.95 \columnwidth]{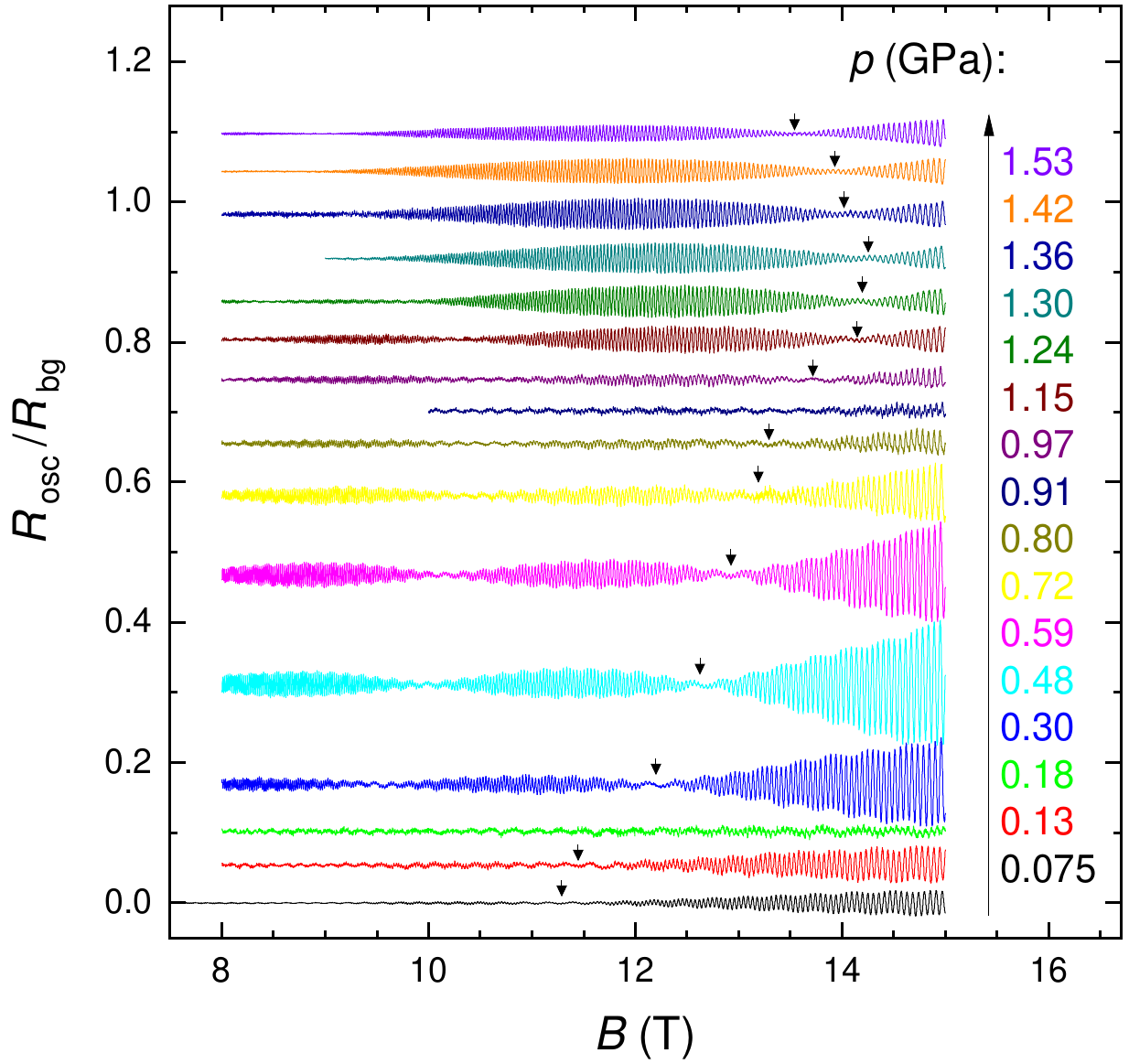}
\caption{SdH oscillations of the interlayer resistance of $\kappa$-Cl, sample \#1 at different pressures, at $T = 0.38$\,K. The oscillating signal is normalized by the nonosillating resistance background, see text. The field is perpendicular to the conducting layers.
The curves are vertically shifted for clarity. Arrows point to the $p$-dependent positions of the higher-field node in the beating $\beta$ oscillations.}
\label{SdH_P}
\end{figure}
\vspace{5pt}
Figure\,\ref{SdH_P} shows examples of the SdH oscillations recorded for $\kappa$-Cl sample \#1 at different pressures, at the base temperature of the $^3$He cryostat.
For each pressure the oscillating signal is normalized by the respective field-dependent nonoscillating background $R_{\text{bg}}$ obtained by a low-order polynomial fit of the as-measured resistance $R(B)$: $R_{\text{osc}}/R_{\text{bg}} \equiv \left[ R(B) - R_{\text{bg}}(B) \right]/R_{\text{bg}}(B)$. In full agreement with the previous reports \cite{kart95c,yama96,ober23},
two fundamental frequencies are observed, revealing the Fermi surface topology typical of the $\kappa$ salts \cite{kart04,sing00,wosn96}, see also Fig.\,\ref{tt}(b) below. The lower frequency, $F_{\alpha}$, is associated with the classical orbit on the Fermi pocket centered at the Brillouin zone boundary and varies between $\approx 530$ and 675\,T upon increasing pressure from 75\,MPa to 1.5\,GPa. The dominant oscillations clearly resolved at all pressures have a higher frequency, $F_{\beta}\sim 4$\,kT, corresponding to a cyclotron orbit area equal to that of the first Brillouin zone. This orbit is caused by magnetic breakdown between the $\alpha$ pocket and a pair of open sheets and thus represents the entire Fermi surface (in the two-dimensional, 2D, representation, i.e., the Fermi surface in the plane of the conducting molecular layers)  \cite{Comment_MB}.

The $\beta$ oscillations exhibit pronounced beating, indicating that there are in fact two frequencies close to each other.
This frequency splitting, $\Delta F_{\beta}/F_{\beta} \sim 0.01$, most likely originates from the maximal and minimal cross-sections of the three-dimensional (3D) Fermi surface cylinder slightly warped in the interlayer direction (see the Supplemental Material of ref. \cite{ober23}), a phenomenon observed earlier on a number of organic \cite{kart88b,kang89,weis99a,schi00} and inorganic \cite{berg00,carr11,seba15,oliv22,eato24} layered materials.
For example, at $p = 0.3$\,GPa we find two beat nodes, at 9.50\,T and at 12.21\,T, respectively. From this we roughly evaluate the warping of the Fermi surface: $\Delta k_{\text{F}}/k_{\text{F}} \simeq \Delta F_{\beta}/2F_{\beta} \approx 0.55 \times 10^{-2}$ (here we used the experimental value $F_{\beta}(0.3\text{GPa})= 3900$\,T) \cite{comm_PhSh}. This warping is similar to that in the isostructural ambient-pressure superconductor $\kappa$-(BEDT-TTF)$_2$Cu[N(CN)$_2$]Br \cite{weis99a} and an order of magnitude stronger than in the sibling $\kappa$-NCS salt \cite{sing02}.

At increasing pressure up to 1.36\,GPa, the beat nodes shift to higher field, indicating an increase of the beat frequency, hence of the Fermi surface warping by $\simeq 25\%$.
A noticeable pressure-induced enhancement of the interlayer coupling is common for the layered organic conductors with their relatively soft crystal lattices.
Interestingly, however, the node positions start to shift down upon further pressurizing beyond 1.4\,GPa.
This apparent weakening of the interlayer coupling at high pressures is unusual and may deserve further attention.

\begin{figure}[tb]
\center
\includegraphics[width = 0.95 \columnwidth]{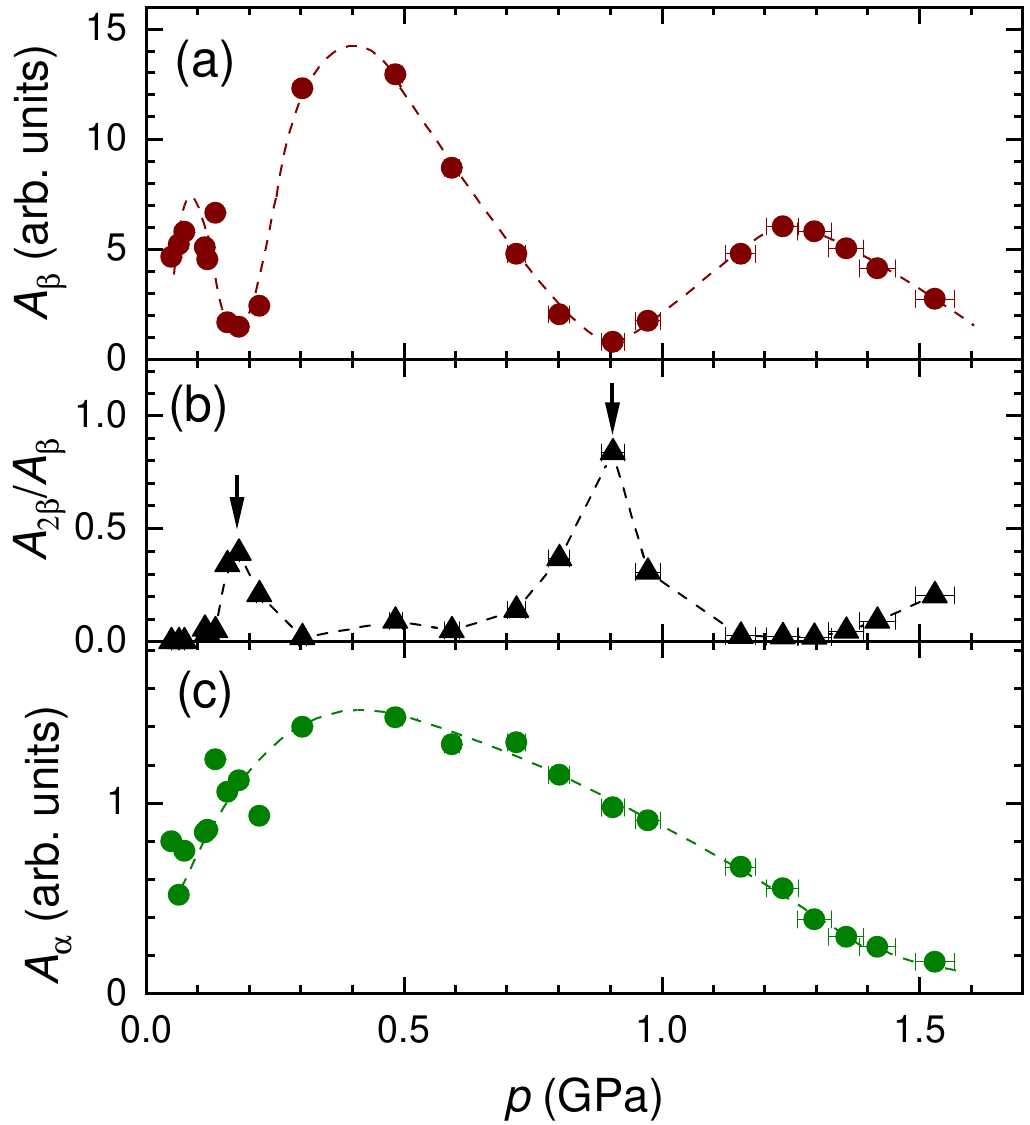}
\caption{Pressure dependence of the FFT amplitudes of the SdH oscillations in $\kappa$-Cl sample \#1 measured in the field window 12 to 15\,T, at $T = 0.38$\,K: (a) amplitude of the $\beta$ oscillations; (b) $2^{\text{\underline{nd}}}$\,-\,to\,-\,$1^{\text{\underline{st}}}$\,-harmonic ratio for the $\beta$ oscillations; (c) amplitude of the $\alpha$ oscillations. Lines are guides to the eye.}
\label{A_P}
\end{figure}

One can see that the amplitude of the oscillations in Fig.\,\ref{SdH_P} varies in a nonmonotonic manner at changing pressure. In Fig.\,\ref{A_P} we plot the $p$-dependence of the amplitudes of main peaks in the fast Fourier transform (FFT) spectra of the oscillatory magnetoresistance, in the field window  12\,T to 15\,T.
The amplitude of the $\beta$ oscillations, see Fig.\,\ref{A_P}(a), exhibits pronounced minima at $p \simeq 0.2$\,GPa and 0.9\,GPa. Simultaneously, the amplitude ratio between the second and first harmonics, $A_{2\beta}/F_{\beta}$, shown in Fig.\,\ref{A_P}(b), displays sharp peaks at the same pressures. This behavior is strongly suggestive of the spin-zero effect caused by the periodic modulation of the oscillation amplitude by the spin-splitting factor $R_{\text{s}}^{(r)} = \cos\left(r\frac{\pi}{2}\frac{m_c}{m_0}g\right)$, where $r$ is the harmonic index, $g$ is the Land\'{e} $g$-factor averaged over the cyclotron orbit, and $m_0$ the free electron mass \cite{shoe84}. For the quasi-2D organics, this effect has been widely known \cite{wosn92,kova93,meye95,weis99b,sasa99,wosn08} as a periodic vanishing of the fundamental harmonic amplitude (with a simultaneously peaking second harmonic) when rotating the magnetic field, due to the angle-dependent cyclotron mass $m_c(\theta) = m_c(0)/\cos \theta$. Knowing the effective cyclotron mass, such angle-dependent data can give useful information about the many-body renormalization of the $g$-factor. It would be interesting to carry out similar measurements on the $\kappa$-Cl salt at different pressures. This should provide data for a comparison between the $p$-dependent renormalization effects on the $g$-factor and on the effective mass.

The variation of the $\alpha$-oscillation amplitude with pressure is shown in Fig.\,\ref{A_P}(c). Here we do not see spin-zero dips. A likely reason for that is a significantly lower cyclotron mass, $m_{c,\alpha} \approx m_{c,\beta}/2$ (see Sect.\,\ref{mass}), which enters the argument of cosine in the expression for $R_{\text{s}}$ and thus leads to its weaker variation under pressure. It is possible that the nonmonotonic behavior with a maximum near $p=0.5$\,GPa is caused by the spin-splitting factor slowly changing with pressure. On the other hand, we note that the $A_{\alpha}(p)$ dependence resembles that of $A_{\beta}(p)$ in Fig.\,\ref{A_P}(a) once we ignore the modulation of the latter by the oscillating spin-splitting factor. In both cases the amplitudes display a global maximum near 0.5\,GPa and a general trend to decrease at high pressures. The mechanisms behind this behavior may be common for the $\alpha$ and $\beta$ oscillations. For example, one can speculate that the initial increase of the amplitude at low pressures is related to the rapid decrease of both cyclotron masses and concomitant weakening of temperature and scattering damping effects on the quantum oscillations \cite{shoe84}. The following slow decrease of the amplitude above 0.5\,GPa may come from a pressure-induced enhancement of the Fermi surface warping (i.e. enhancement of the interlayer coupling), which should lead to a decrease of the number of charge carriers contributing in phase to the quantum oscillations.

\begin{figure}[tb]
\center
\includegraphics[width = 0.95 \columnwidth]{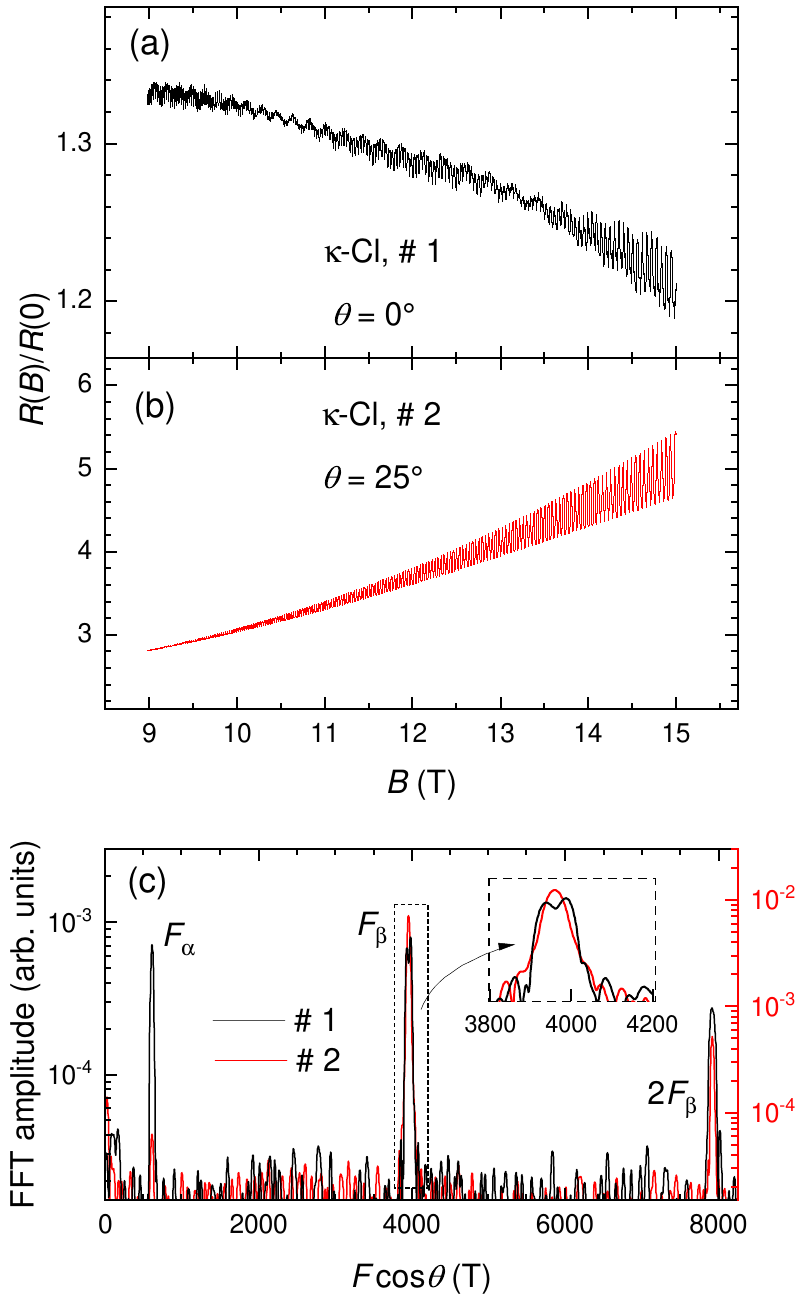}
\caption{Field-dependent resistance of  two $\kappa$-Cl samples aligned differently with respect to the magnetic field and measured simultaneously at $p = 0.8$\,GPa, $T = 0.38$\,K. (a) Sample \#1, the field perpendicular to layers; (b) Sample \#2 with the field directed near the first AMRO peak (see text). (c) FFT of the oscillating signal for sample \#1 (black, left Y-axis) and sample \#2 (red, right Y-axis). Inset: close-up of the $F_{\beta}$ peaks for both samples. For sample \#1 the peak is split, in line with the beating behavior of the oscillations.}
\label{SdH_ex}
\end{figure}
The beating of the $\beta$ oscillations and the resulting splitting of the FFT peak obviously affects the precision of determination of the mean Fermi surface area, which we will need in the following for evaluation of the frustration parameter $t'/t$.
This source of error can be avoided by aligning the sample in the direction corresponding to a maximum in the classical angle-dependent magnetoresistance oscillations (AMRO) \cite{kart88b,yama89,kart90,wosn93,pesc02,kart08a}. At such directions, known also as Yamaji angles, all cyclotron orbits on a weakly warped cylindrical Fermi surface have the same area \cite{yama89}, hence, contribute in phase to the quantum oscillations. As a result, when the sample is turned in a magnetic field, approaching a Yamaji angle, the beat frequency vanishes and the SdH amplitude acquires a local maximum.
This effect is illustrated in Fig.\,\ref{SdH_ex}, where the field-dependent resistance of two samples mounted side by side in the pressure cell but aligned differently with respect to the magnetic field is plotted. Here, panel (a) shows sample\,\#1 aligned with its conducting layers perpendicular to the field. Sample\,\# 2 in Fig.\,\ref{SdH_ex}(b) is tilted from the perpendicular orientation by angle $\theta \approx 25^{\circ}$, which is close to the first Yamaji angle for the $\beta$ orbit \cite{yama96}. As a result, the amplitude of the $\beta$ oscillations is strongly enhanced and no trace of beating is seen. The fast Fourier transforms (FFT) of both oscillating signals are shown in  Fig.\,\,\ref{SdH_ex}(c). Here, in order to facilitate a direct comparison between different orientations, we multiply the frequencies by $\cos \theta$, thereby reducing them to the values corresponding to $\theta = 0^{\circ}$.
One readily sees that the $F_{\beta}$ peak for sample \#2 (red curve) is greatly enhanced and, in contrast to sample \#1 (black curve), shows no splitting.

Thus, at $\theta \approx 25^{\circ}$  we significantly gain in the accuracy of both the frequency and amplitude of the $\beta$ oscillations. Therefore, most of the measurements on sample \#2 was done at this orientation. Note, however, that at this orientation the information on the Fermi surface warping is lost and the $\alpha$ oscillations are significantly suppressed, in comparison with those in the perpendicular field. Hence, a part of the measurements on sample \#2 and all studies of sample \#1 were done at $\theta = 0^{\circ}$.
In what follows, we will present the SdH frequencies and effective masses obtained for $\kappa$-Cl in both the perpendicular and the tilted orientations.


We also note that due to the very high electronic anisotropy of the $\kappa$-NCS salt (see the Supplemental Material \cite{sm-hp} and references therein) the abovementioned effect of AMRO on the SdH oscillations is absent in this material. Therefore, all measurements on $\kappa$-NCS have been done in the perpendicular field geometry.

\subsection{$p$-dependent SdH frequencies and inplane anisotropy} \label{freq}
\begin{figure}[tb]
\center
\includegraphics[width = 1 \columnwidth]{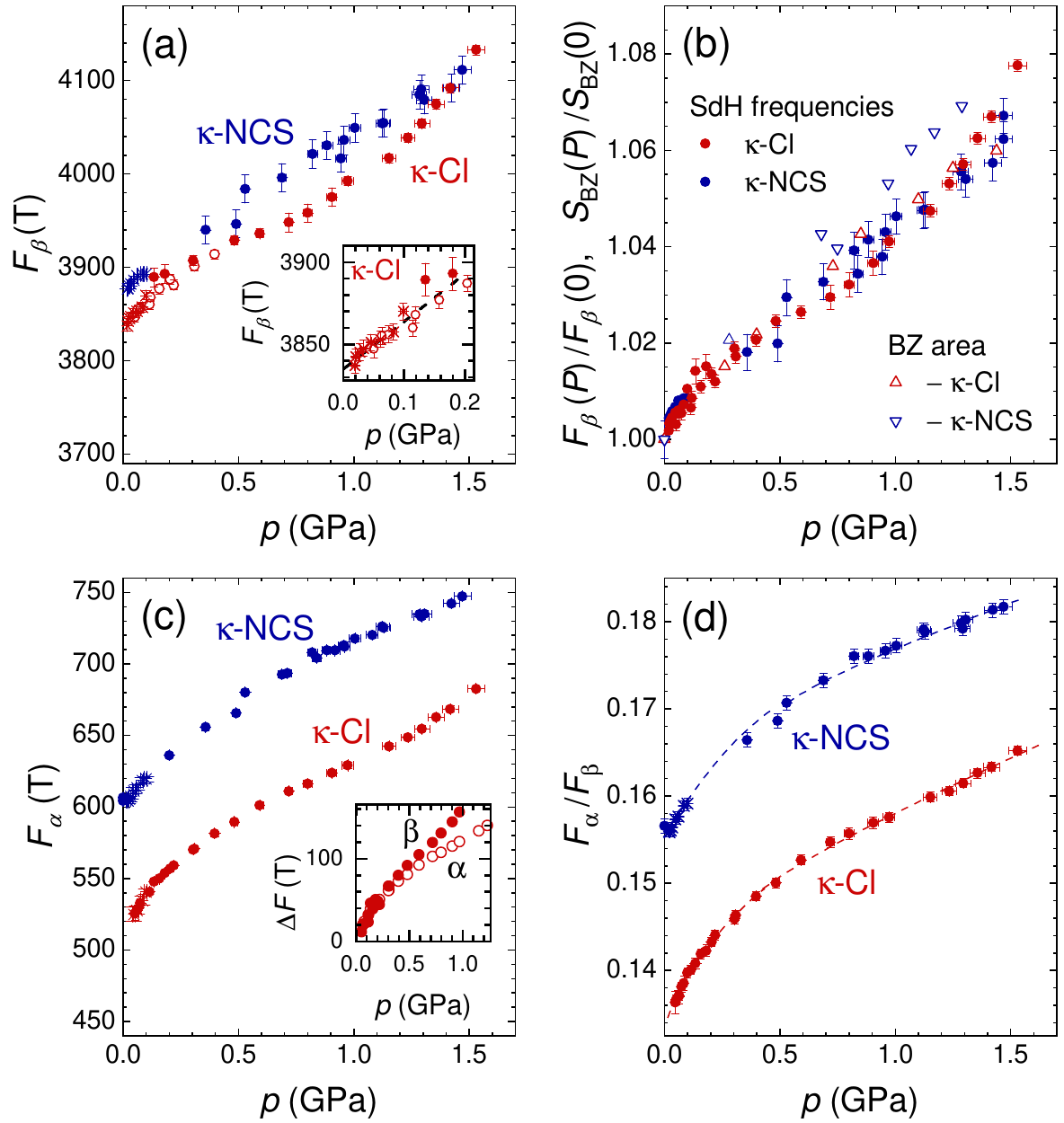}
\caption{Pressure-dependent SdH frequencies in  $\kappa$-Cl (red) and in $\kappa$-NCS (blue). Different symbols represent different samples or field directions; see text for explanation. (a) $\beta$-oscillations.
Inset: close-up of the lower-pressure range for $\kappa$-Cl. The dashed line is a linear fit to the data, yielding the zero-pressure extrapolation $F_{\beta}(0) \approx 3840$\,T.
(b) The data from panel (a) normalized to the respective zero-pressure frequency (circles). For simplicity, no distinction between different samples is made. For comparison, the relative change of the $p$-dependent Brillouin zone area $S_{\text{BZ}}(p)/S_{\text{BZ}}(0)$
is shown by triangles.
(c) $\alpha$-oscillations. Inset: absolute changes of the $\alpha$ (open circles) and $\beta$ (filled circles) frequencies in $\kappa$-Cl under pressure: $\Delta F \equiv F(p)-F(0)$. (d) Pressure dependence of the ratio $F_{\alpha}/F_{\beta}.$
}
\label{F_P}
\end{figure}

In this section we present a detailed analysis of the SdH frequencies, which are fundamentally determined by the area of the relevant Fermi surface cross-section $S_i$ through the Onsager relation \cite{shoe84}, $F_i = \frac{\hbar e}{2\pi}S_i$, with $\hbar$ being the Planck constant and $e$ the elementary charge.
Figure \ref{F_P}(a) shows the pressure-dependent frequencies of the $\beta$ oscillations in $\kappa$-Cl sample \#2 (red symbols)  and $\kappa$-NCS (blue symbols).
For $\kappa$-Cl, the empty circles correspond to the perpendicular field geometry and the filled circles are the data taken at $\theta = 25^{\circ}$  and multiplied by $\cos \theta$. Sample \#1 was measured simultaneously with \#2, but at $\theta = 0^{\circ}$ and showed consistent values within the error bars. For $\kappa$-NCS, the filled circles are the averaged values obtained on two samples  measured simultaneously; the difference between the samples lies within the indicated error bars.
The stars are the data obtained in our previous dilution-fridge experiment on $\kappa$-Cl and $\kappa$-NCS at pressures up to 0.1\,GPa \cite{ober23}.

 As already mentioned, the $\beta$ oscillations are associated with the magnetic-breakdown orbit with the area equal to the Brillouin zone area.
 Indeed, the zero pressure values, $F_{\beta}^{\text{Cl}}(0) = (3836 \pm 5)$\,T and $F_{\beta}^{\text{NCS}}(0) = (3867 \pm 7)$\,T yield the areas $(36.62\pm 0.05)$\,nm$^{-2}$ and $(36.91\pm 0.05)$\,nm$^{-2}$, respectively,
perfectly coinciding with the low-temperature Brillouin zone areas of these salts \cite{wata99,schu91,comm_F0}.

The shapes of the $p$ dependence in Fig.\,\ref{F_P}(a) look slightly different from each other. However, by plotting the relative change of the frequency under pressure, see Fig.\,\ref{F_P}(b), we find that the difference between the two salts does not exceed the experimental error bars. Moreover, our data are consistent with the quasi-linear $p$ dependence of the Brillouin zone areas [triangles in Fig.\,\ref{F_P}(b)] based on the X-ray data \cite{schu94,raha97}. We note that the X-ray studies  \cite{schu94,raha97} have been done at room temperature. However, their good agreement with our low-temperature SdH data suggests that the compressibility does not change significantly upon cooling.

Plotted in Fig.\,\ref{F_P}(c) is the pressure dependence of the $\alpha$ frequency. For $\kappa$-NCS the symbol and color codes are the same as in Fig.\,\ref{F_P}(a). For $\kappa$-Cl, the data (circles) have been taken on sample \#1 in the perpendicular field geometry. For pressures below 0.1\,GPa, the results from our previous dilution fridge experiment \cite{ober23} are added (stars). Both data sets are perfectly consistent with each other. Therefore, we will not distinguish between them in the following.

For $\kappa$-NCS our data set is consistent with the early study by Caulfield et al. \cite{caul94}. As was already noticed by those authors, the relative increase of $F_{\alpha}$ with pressure is much stronger than that of $F_{\beta}$. The $\kappa$-Cl salt shows the same, even more pronounced trend.
At lower pressures, $p < 0.3$\,GPa, the $\alpha$ frequency increases at a relative rate of  $\simeq 0.25\,\text{GPa}^{-1}$ for $\kappa$-NCS and $\simeq 0.4\,\text{GPa}^{-1}$ for $\kappa$-Cl [cf. the relative increase rate of the $\beta$ frequencies in Fig.\,\ref{F_P}(b) is only $\approx 0.04\,\text{GPa}^{-1}$].
Interestingly, for $\kappa$-Cl, the absolute changes of $F_{\alpha}(p)$ and $F_{\beta}(p)$ are virtually the same in this pressure range, see the inset in Fig.\,\ref{F_P}(c). This suggests that, in contrast to the rapidly expanding $\alpha$ pocket, the rest of the Fermi surface remains almost unchanged. At higher pressures, the increase of $F_{\alpha}$ becomes more moderate, with a slope saturating at $\sim 0.15\,\text{GPa}^{-1}$.
For $\kappa$-NCS, due to the weakness of the $\beta$ oscillations at $p< 0.3$\,GPa,
a sufficiently accurate comparison of $\Delta F_{\alpha}(p)$ and $\Delta F_{\beta}(p)$ is difficult. However, qualitatively, the behavior is similar to that of $\kappa$-Cl.

The difference in the behaviors of the $\alpha$ and $\beta$ frequencies is summarized as the $p$-dependent ratio $F_{\alpha}/F_{\beta}$ in Fig.\,\ref{F_P}(d).
Let us discuss this ratio in terms of the electronic anisotropy of the conducting layers, in other words, in terms of the shape of the 2D Fermi surface.
To this end, we follow the approach \cite{caul94,prat10b} based on the effective dimer model commongly used for the $\kappa$ salts. This is a tight-binding model of an anisotropic triangular lattice of BEDT-TTF dimers with the nearest and next-nearest transfer integrals, $t$ and $t^{\prime}$, respectively [see inset in Fig.\,\ref{tt}(a)], and the dispersion relation:
\begin{figure}[tb]
\center
\includegraphics[width = 0.95 \columnwidth]{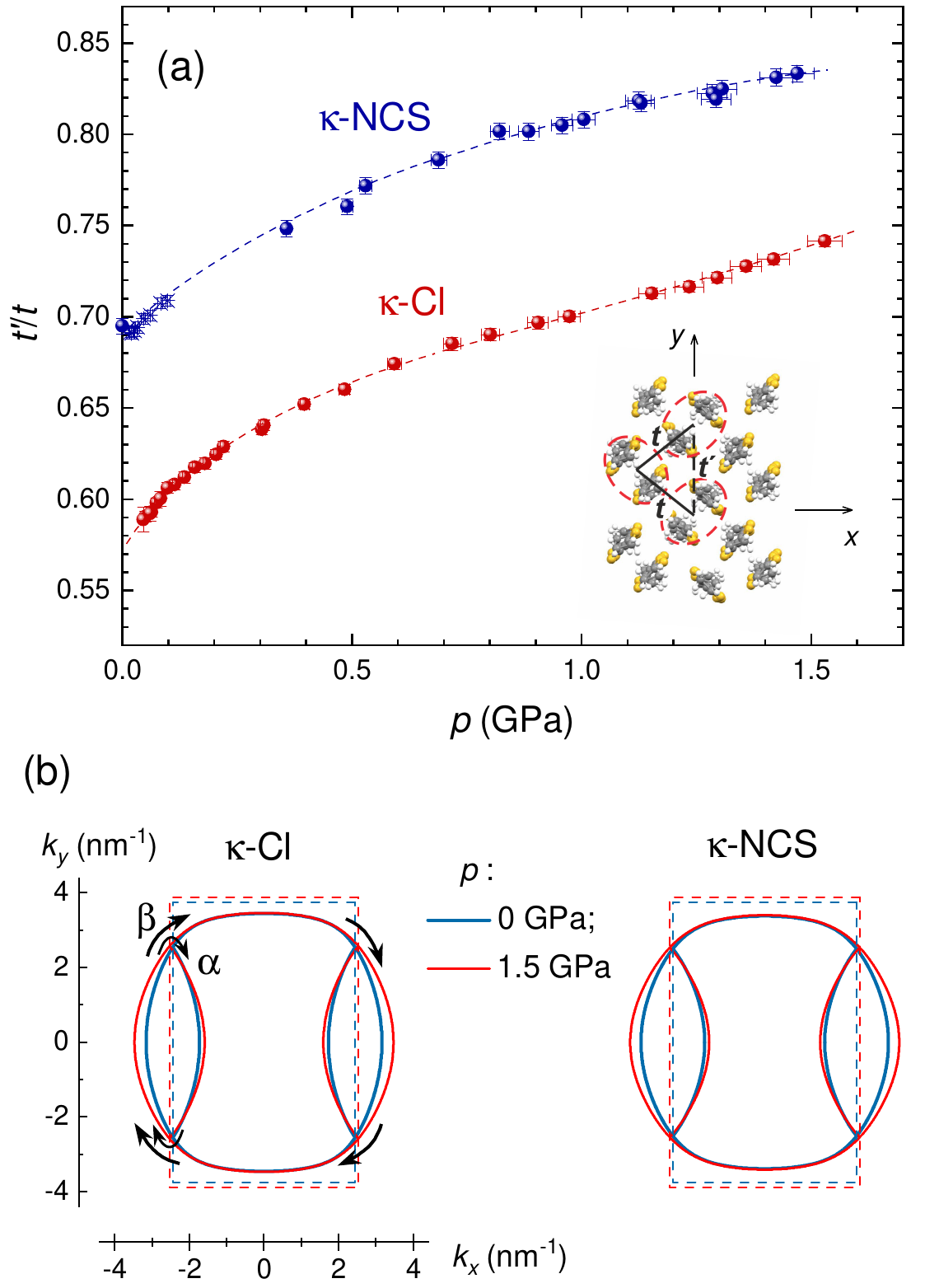}
\caption{(a) Spin frustration ratio $t'/t$ in $\kappa$-Cl (red) and in $\kappa$-NCS (blue) calculated from the data in Fig.\,\ref{F_P}(d) using Eq.\,(\ref{tbm}). Inset: fragment of the molecular layer viewed along the BEDT-TTF long axis. The molecules are arranged in dimers (dashed red lines) forming an anisotropic triangular lattice with the transfer integrals $t$ and $t'$, schematically shown by the thin solid and dashed lines, respectively.
(b) Solid lines: Fermi surfaces of $\kappa$-Cl and $\kappa$-NCS calculated, within the effective dimer model, for ambient pressure and for $p=1.5$\,GPa. Dashed lines: respective Brillouin zone boundaries. The linear scales were calculated based on the low-$T$ crystallographic parameters \cite{wata99,schu91} and room-temperature compressibilities  \cite{schu94,raha97}. The arrows show schematically the Bragg reflections and tunneling at the magnetic-breakdown junctions leading, respectively, to the $\alpha$ and $\beta$ orbits in strong magnetic fields.}
\label{tt}
\end{figure}
\begin{equation} \label{tbm}
\epsilon(\mathbf{k}) = 4t\cos\left( \frac{k_x x}{2} \right) \cos\left( \frac{k_y y}{2} \right) +2t^{\prime} \cos \left( k_y y \right).
\end{equation}
Here, $x$ and $y$ should be substituted by the crystallographic parameters $a$($c$) and $c$($b$) in $\kappa$-Cl(-NCS), respectively.
The above equation is a parametric expression for the Fermi surface which directly determines the ratio  between the Fermi surface areas $S_{\alpha}$ and $S_{\beta}$, hence the ratio $F_{\alpha}/F_{\beta}$ through $t^{\prime}/t$, the spin frustration parameter of the anisotropic triangular lattice of the molecular dimers \cite{prat10b}.

Fitting the experimental data in Fig.\ref{F_P}(d) with Eq.\,(\ref{tbm}), we evaluate the frustration ratio $t^{\prime}/t$  and its dependence on pressure for both salts, as shown in Fig.\,\ref{tt}(a).
The first, obvious result is that the frustration in the metallic $\kappa$-NCS salt  is significantly stronger than in the ambient-pressure Mott insulator $\kappa$-Cl. While this difference was predicted by some band structure calculations \cite{mori99a,kand09,kore14}, our data provide, to the best of our knowledge, the first direct experimental evidence for that.

Another important observation is that even our quasi-hydrostatic pressure significantly changes the electronic anisotropy in the conducting layers, leading to an enhancement of the spin frustration.
In a broad pressure range the $t^{\prime}/t$ ratio increases with an approximately constant rate of $\simeq 0.07\,\text{GPa}^{-1}$, which even increases below 0.3\,GPa, as the system approaches the MIT. The overall increase of the frustration ratio in the studied pressure range is $\gtrsim 20\%$. In particular, at 1\,GPa the frustration in the $\kappa$-Cl salt already exceeds the ambient-pressure value for $\kappa$-NCS.
Thus, besides the well-known effect of pressure on electronic correlation strength, it is important to take into account its strong influence on magnetic ordering instability in these materials.

In Fig.\,\ref{tt}(b) we show the Fermi surfaces of $\kappa$-Cl and $\kappa$-NCS, calculated using Eq.\,(\ref{tbm}) and the experimental SdH frequencies, for the lowest and for the highest pressure. Even though the crystal lattice compressibility is assumed to be isotropic in the layers plane, as it is at room temperature \cite{schu94,raha97}, the changes in the Fermi surfaces of both salts are obviously anisotropic. While the $\alpha$ pocket shows a significant increase along its short axis ($\| k_x$), the rest of the Fermi surface remains almost the same. Such a behavior is indeed observed experimentally on $\kappa$-Cl at pressures of up to $0.3$\,GPa, as noted above. At higher pressures, however, the absolute changes $\Delta F_{\beta}(p)$ and $\Delta F_{\alpha}(p)$ deviate from each other, see, e.g., the inset in Fig.\,\ref{F_P}(c). This means that the Fermi surface also expands in the $k_y$ direction. The apparent absence of such expansion in Fig.\,\ref{tt}(b) is most likely due to a limited precision of the simple effective dimer model employed here. It would be highly interesting to perform a more elaborated analysis confronting our data with an ab-initio band structure calculation taking into account electronic correlations. This, however, appears to be a very challenging task, requiring, furthermore, detailed low-$T$ structural data at high pressures. On the other hand, our main conclusions concerning the comparison of the spin frustration in the present two compounds and their pressure dependence seem to be  robust against small quantitative corrections.

\subsection{Effective cyclotron masses} \label{mass}
The effective cyclotron masses were evaluated from the $T$-dependence of the amplitude $A_i(T)$ of the fundamental harmonic of the SdH oscillations in a conventional way based on the Lifshitz-Kosevich (LK) theory \cite{lifs55,shoe84}.
Details of the evaluation including some examples are given in the Supplemental Material \cite{sm-hp}.


\begin{figure}[tb]
\center
\includegraphics[width = 0.95 \columnwidth]{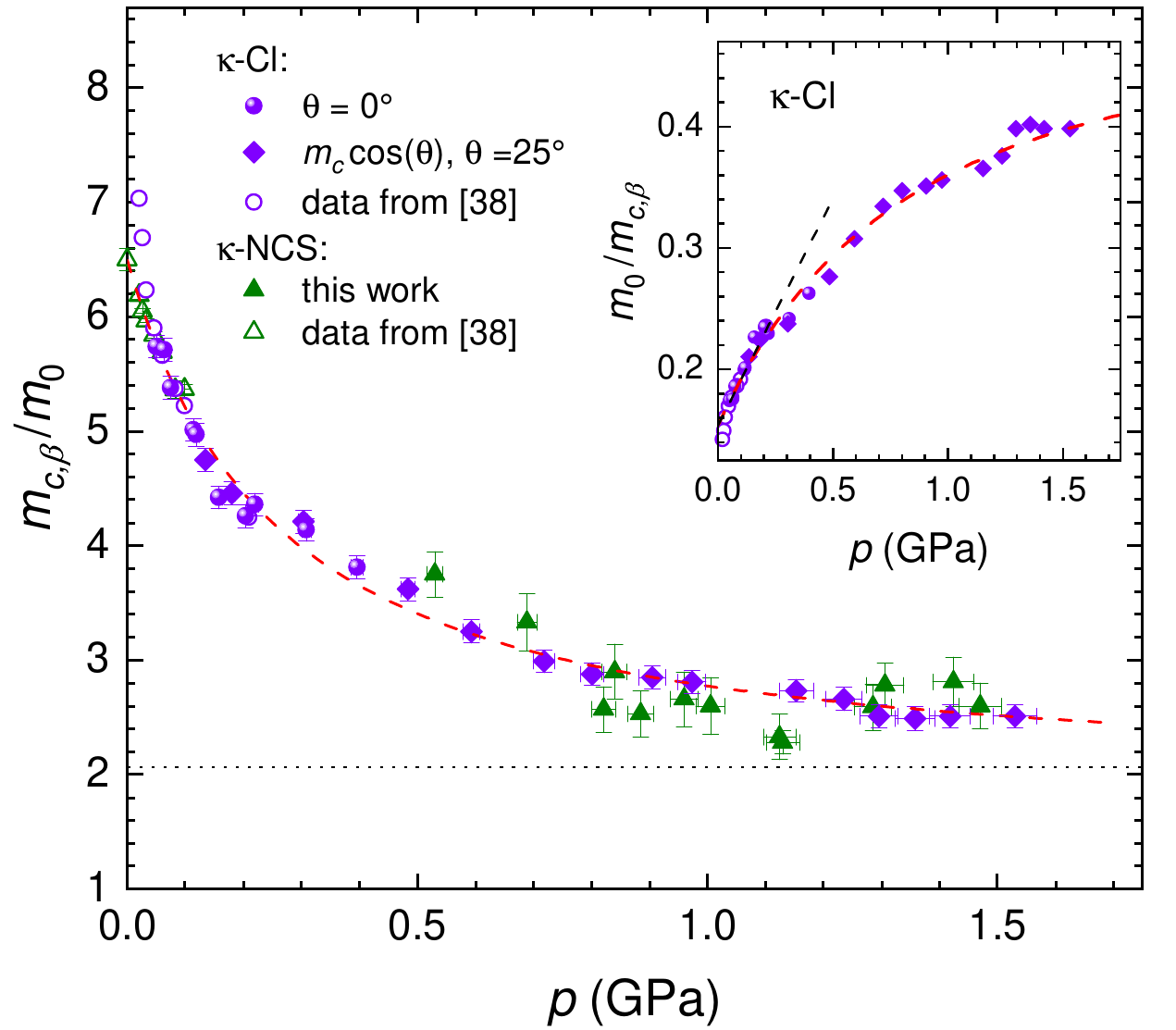}
\caption{Pressure dependence of the effective cyclotron mass on the $\beta$ orbit for $\kappa$-Cl (circles and diamonds) and $\kappa$-NCS (triangles). The data is normalized by the free electron mass $m_0$. Results from the present experiment and from the lower-pressure study \cite{ober23} are shown. The red dashed line is a fit to the $\kappa$-Cl data using Eq.\,(\ref{mBR}). The horizontal dotted line indicates the level of the noninteracting band cyclotron mass $m_{c,\beta,
\text{band}}$ obtained from the fit. Inset: the inverse mass of $\kappa$-Cl versus pressure. The red line shows the same fit as in the main panel. The black straight line is the linear fit from \cite{ober23}.}
\label{m_P}
\end{figure}
The results for the mass (in units of the free electron mass $m_0$) on the $\beta$ orbit, characterizing the entire Fermi surface, are plotted in Fig.\,\ref{m_P}. Here the blue  symbols represent $\kappa$-Cl sample \#2. The filled circles and diamonds correspond to the data obtained at $\theta = 0^{\circ}$ (field perpendicular to the layers) and at $25^{\circ}$ (near the AMRO peak), respectively. The latter are multiplied by $\cos 25^{\circ}$.
Sample \#1, measured simultaneously with sample \#2, yielded very similar mass values (not shown in Fig.\,\ref{m_P}, for the sake of clarity). The empty circles are the data obtained on sample \#1 in our earlier lower-pressure experiment \cite{ober23}.
The green symbols show the mass for the $\kappa$-NCS salt obtained in this work (filled triangles) and in Ref. \cite{ober23} (empty triangles).
Due to a larger magnetic breakdown gap, the $\beta$ oscillations in $\kappa$-NCS are relatively weak (see the Supplemetal Material \cite{sm-hp}), which leads to a larger error bar and stronger scattering of the data.

Within the experimental error, the $\kappa$-Cl and $\kappa$-NCS salts exhibit the same behavior. The initial rapid decrease of the mass occurring as
we are moving away from the MIT slows down with increasing pressure and saturates above 1\,GPa at the level $m_{c,\beta}\simeq 2.5 m_0$. Note that this value is close to the band cyclotron mass $m_{c,\beta,\text{band}} = 2.6m_0$ \cite{meri00a} calculated for $\kappa$-NCS from the band structure neglecting many-body interactions. Thus, it appears that at pressures above 1\,GPa we have essentially a noninteracting electron system. Taking into account that the only significant change in our materials within the rather narrow range $0 < p < 1$\,GPa is the electronically driven MIT, it is natural to attribute the effective mass renormalization entirely to electronic correlations.
Other renormalization effects, such as due to electron-phonon interactions, not directly linked to the MIT, are not expected to change notably within this pressure range and, therefore, are most likely insignificant.

The above argument justifies the analysis of the pressure-dependent effective mass in terms of the electronic correlation strength ratio $U/t$. In particular, the fact that the masses $m_{c,\beta}$ in $\kappa$-Cl and $\kappa$-NCS are very similar at all pressures and approach the same high-$p$ limit $m_{c,\beta,\text{band}}$ further substantiates the earlier conclusion \cite{ober23}, inferred from lower-pressure data, that the correlation strength is indeed essentially the same in both salts.

Turning to a more quantitative analysis, the previous, low-pressure experiment \cite{ober23} has revealed a simple inverse-linear $p$-dependence of the mass, which was interpreted in terms of a Brinkman-Rice-like (BR-like) renormalization \cite{brin70,geor96}. The present data, obtained in a broader pressure range, reveal a deviation from this behavior starting from at $p \simeq 0.25$\,GPa.
This is clearly seen in the inset in Fig.\,\ref{m_P}, where the inverse mass of $\kappa$-Cl is plotted and the dashed straight line is the linear fit to the low-pressure data \cite{ober23}. It should be noted that the linear dependence, $m_c^{-1} \propto (p-p_0)$, was inferred from the BR theory \cite{brin70}, assuming a linear pressure dependence of the correlation strength ratio. This approximation works well in a narrow pressure interval near the MIT, where the change of the ratio $U/t$ does not exceed $1-2\%$ \cite{ober23}.
However, for a broader range one should take into account that the inter-site transfer integral $t$ is significantly more sensitive to pressure than the on-site (intra-dimer) Coulomb repulsion $U$ \cite{mori99a}.
In this case, a more reasonable approximation is \cite{seme23} to assume $U = \text{const}$ and expand $t$ rather than $U/t$ linearly in pressure:
\begin{equation}\label{texp}
t(p) \approx t_0\left[1 + \gamma(p-p_0)\right],
\end{equation}
where $p_0$ is the critical pressure where the mass diverges in the BR model, $t_0 \equiv t(p_0)$ and $\gamma$ is a proportionality factor. Further following  \cite{seme23}, we set the critical on-site repulsion $U_0$ proportional to the conducting bandwidth, hence to $t(p)$. Then, the BR renormalization of the effective mass \cite{brin70}, $m_c =  \frac{m_{c,\text{band}}}{\left[1 - \left(U/U_0\right)^2  \right]}$, can be as:
\begin{equation}\label{mBR}
m_c =
 m_{c,\text{band}}\left[1-\frac{1}{\left[1+\gamma(p-p_0)\right]^2}\right]^{-1}\,.
\end{equation}
Here, we replaced the usual quasiparticle effective mass $m^{\ast}$ and band mass $m_{\text{band}}$ considered in the original BR theory \cite{brin70} by the respective cyclotron masses considered in the LK theory of magnetic quantum oscillations, since the many-body renormalization effects are the same in both cases \cite{shoe84}.

Despite its rather simple form,  Eq.\,(\ref{mBR}) fits the experimental data remarkably well throughout the whole pressure range. This is shown by the red dashed line in Fig.\,\ref{m_P} fitting the $\kappa$-Cl data \cite{comm_rang}.
We, therefore, assume that it provides a realistic estimate of the parameters characterizing the electronic system: $m_{c,\beta,\text{band}} = (2.07 \pm 0.1)m_0$, $p_0 = (-0.28 \pm 0.04)$\,GPa, and $\gamma = (0.77 \pm 0.11)\,\text{GPa}^{-1}$. The fit to the $\kappa$-NCS dataset yields very similar parameters, although with considerably larger error bars, see the Supplemental Material \cite{sm-hp}. The evaluated band mass is somewhat smaller than the abovementioned calculated value, $2.6m_0$ \cite{meri00a}, but the difference does not exceed the uncertainty of the band structure calculations \cite{Comment_mass}. The BR critical pressure is comparable to that obtained from an even simpler  low-$p$ fit \cite{ober23} shown by the dashed straight line in the inset in Fig.\,\ref{m_P}. Finally, within the present approach, the sensitivity of the electronic correlation strength to pressure is basically determined by the coefficient $\gamma = \frac{dt/t_0}{dp}$.
The obtained value is an order of magnitude higher than that inferred from the band-structure calculations \cite{kand09}. The calculations yielded the $U/t$ ratio in $\kappa$-NCS decreasing by only $5\%$ upon increasing pressure from 0 to 0.75\,GPa, which would imply $\gamma \approx 0.07$\,GPa$^{-1}$. A similarly strong disagreement with the theoretical predictions has already been detected in the experimental data taken in a narrow pressure interval very close to the MIT \cite{ober23}. Now it is confirmed to exist over a much broader range where the effective mass is no longer inversely-linear in $p$ and even approaches the noninteracting band mass value.


\begin{figure}[tb]
\center
\includegraphics[width = 0.95 \columnwidth]{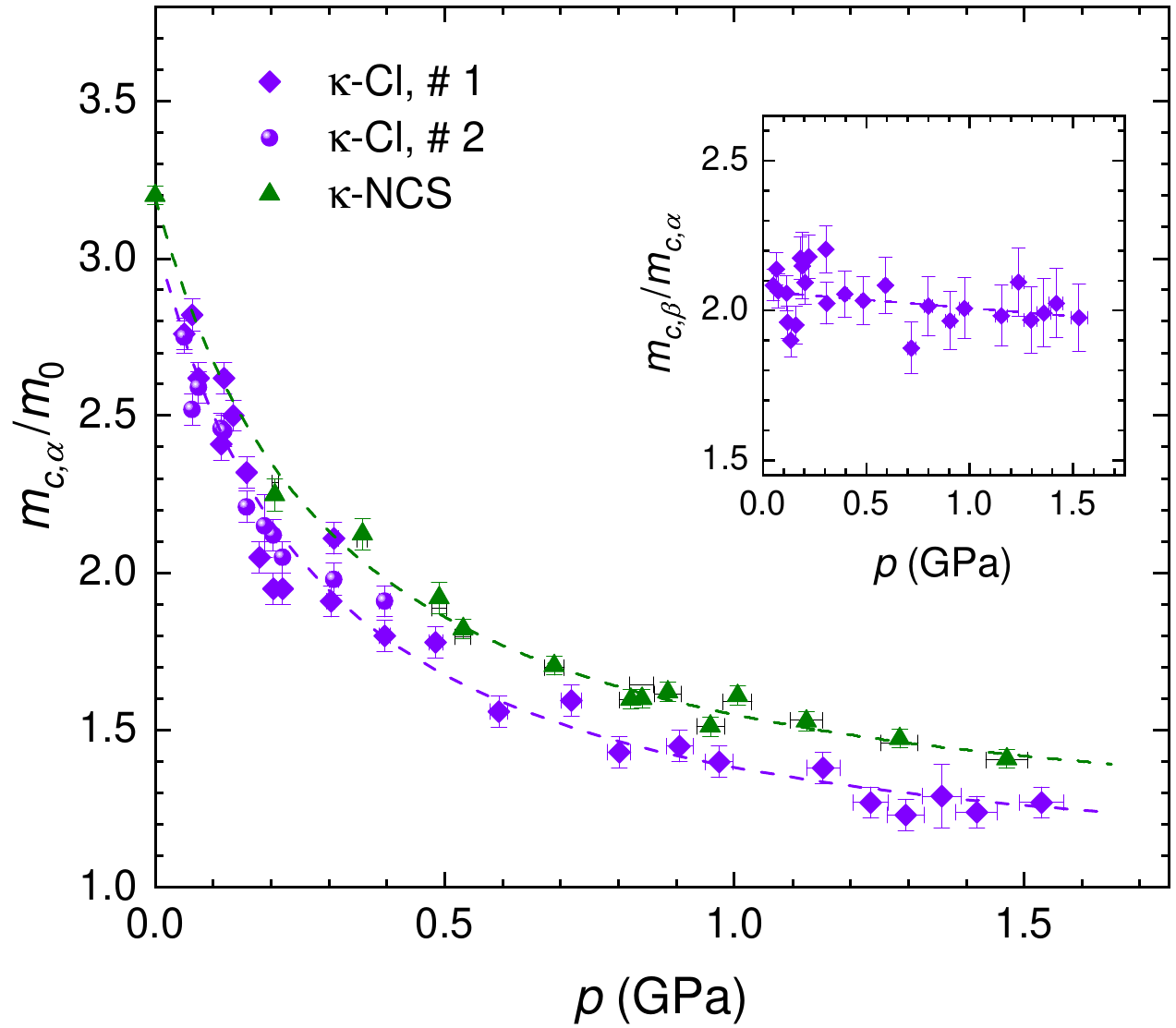}
\caption{Pressure dependence of the effective cyclotron mass on the $\alpha$ orbit for $\kappa$-Cl (circles and diamonds) and $\kappa$-NCS (triangles). 
 The dashed lines are fits with Eq.\,(\ref{mBR}).  Inset: the ratio of the $\beta$- and $\alpha$-masses for $\kappa$-Cl.}
\label{ma_P}
\end{figure}
Thus far, we have considered the cyclotron mass on the magnetic-breakdown $\beta$ orbit encircling the entire 2D Fermi surface. It is interesting to compare it with the mass on the classical orbit $\alpha$, which involves only one-half of the charge carriers. In this way we may obtain information on the momentum dependence of the electronic interactions. For example, in another organic salt, $\kappa\text{-(BETS)}_2\text{Mn[N(CN)}_2]_3$, displaying the MIT, the mass renormalization on the $\alpha$ orbit was found to be enhanced in comparison to the rest of the Fermi surface in close proximity to the transition \cite{zver19}. Figure\,\ref{ma_P} summarizes our  results on $m_{c,\alpha}(p)$ in $\kappa$-Cl and $\kappa$-NCS. All the data in the figure have been taken in the perpendicular field configuration, as at the tilted field the $\alpha$ oscillations in $\kappa$-Cl were too weak (see Fig.\,\ref{SdH_ex}) for an accurate mass determination.

For $\kappa$-Cl, the $\alpha$ mass is almost exactly one-half of the $\beta$ mass and this relation is virtually independent of pressure. To illustrate this, we plot the ratio $m_{c,\beta}/m_{c,\alpha}$ in the inset in Fig.\,\ref{ma_P}. A linear fit to the data (dashed line in the inset), has a very slight slope, $(-0.055 \pm 0.04)$\,GPa$^{-1}$, lying within the error bar. Such a weak variation, even if it reflects a real trend, may be attributed to a weak pressure dependence of the band masses.
The blue dashed line in the main panel of Fig.\,\ref{ma_P} is the BR fit according to Eq.\,(\ref{mBR}). It yields the parameters: $m_{c,\alpha,\text{band}} = (1.05 \pm 0.07)m_0$, $p_0 = (-0.29 \pm 0.04)$\,GPa, and $\gamma = (0.81 \pm 0.16)\,\text{GPa}^{-1}$.
As expected, the band mass is approximately one-half of the $\beta$ band mass obtained above. The other two parameters are very close to those obtained for the $\beta$ orbit. All in all, we observe no evidence of a difference in the mass renormalization on the $\alpha$ and $\beta$ orbits. Thus, within the accuracy of our experiment the electronic correlations appear to be  momentum-independent in the $\kappa$-Cl salt, in contrast to those in $\kappa\text{-(BETS)}_2\text{Mn[N(CN)}_2]_3$.

For the $\kappa$-NCS salt (green triangles in Fig.\,\ref{ma_P}), the $\alpha$-mass values lie slightly higher than for $\kappa$-Cl \cite{comm_caul-}. The difference is likely caused by a larger, than in $\kappa$-Cl, size of the $\alpha$ orbit (see Sec.\,\ref{freq}), hence a higher band cyclotron mass. Indeed, the fit with Eq.\,(\ref{mBR}) gives $m_{c,\alpha,\text{band}} = (1.18 \pm 0.06)m_0$, that is $10\%$ higher than in the $\kappa$-Cl salt.
As to the other fitting parameters, the sensitivity to pressure, $\gamma = (0.78 \pm 0.13)\,\text{GPa}^{-1}$ is almost the same as for $\kappa$-Cl, whereas the  BR critical pressure $p_0 = (-0.33 \pm 0.03)$\,GPa is slightly lower.
It is also lower than the value, $p_0 \simeq -0.29$\,GPa, obtained from the fit to the $m_{c,\beta}(p)$ dependence in the same salt, see Fig.\,S4 in the Supplemental Material \cite{sm-hp}.
The relatively low $p_0$ value for the $\alpha$ orbit might be a sign of weaker electronic correlations. If so, this would imply that the correlation effects are different on different parts of the Fermi surface in $\kappa$-NCS, weaker on the $\alpha$ pocket and stronger on the open sheets. However, we should keep in mind that the mentioned differences in $p_0$ are small and comparable to the evaluation error bars. Here, the limiting factor is the rather large error bars in the $m_{c,\beta}(p)$ dependence caused by the low amplitude of the magnetic-breakdown $\beta$ oscillations in $\kappa$-NCS. For making a definitive conclusion, further measurements at higher magnetic fields, $B>30$\,T, would be very helpful. While high-field quantum oscillations experiments have been done on $\kappa$-NCS at ambient pressure, see, e.g., refs. \cite{audo16,godd04}, we are unaware of similar measurements under pressure.


\section{Conclusions}\label{conc}
Using the SdH oscillation technique, we have been able to trace the evolution of the electronic correlation strength as well as the spin frustration ratio in the $\kappa$-(BEDT-TTF)$_2$X salts with X\,= Cu[N(CN)$_2$]Cl and Cu(NCS)$_2$ in a broad pressure range up to 1.5\,GPa corresponding to an almost two-fold change of the conduction bandwidth, according to our estimations.

From the systematic analysis of the SdH amplitude, we have determined the renormalized
effective cyclotron masses. The renormalization is found to be the same for the $\alpha$ and $\beta$ orbits. This suggests that electronic correlations are homogeneous over the Fermi surface.
Throughout the entire pressure range studied the behavior of the effective cyclotron mass is remarkably well described by the BR model under the assumption of a linear-in-pressure transfer integral $t(p)$ and a $p$-independent on-site Coulomb repulsion. This approximation was recently shown to work well for the inorganic 3D Mott insulator NiS$_2$ at pressures between 3 and 11\,GPa \cite{seme23} and also seems to be very reasonable in our case. The sensitivity of the transfer integral to pressure, $\gamma = \frac{dt/t_0}{dp} \simeq 0.8$\,GPa$^{-1}$, was estimated by fitting the experimental data with the model Eq.\,(\ref{mBR}). This result is consistent with the previous estimations based on low-pressure data \cite{ober23}, but is an order of magnitude higher than inferred from the band structure calculations \cite{kand09}. This stark discrepancy challenges our understanding of the correlation effects on the band structure near the bandwidth-controlled MIT.

Our data confirm and further extend to a broad pressure range the earlier finding \cite{ober23} that the correlation strength is the same in the ambient-pressure insulator $\kappa$-Cl and in the superconducting $\kappa$-NCS. By contrast, the spin frustration turns out to differ considerably in the two salts. To estimate the frustration ratio $t^{\prime}/t$, we use the fact that this parameter is intimately connected with the shape of the Fermi surface and thus can be extracted from the relationship of the SdH frequencies $F_{\alpha}$ and $F_{\beta}$. To this end we followed the approach based on the effective dimer model \cite{caul94,prat10b}. The resulting ambient-pressure values are $t^{\prime}/t \approx 0.57$ and 0.69 for $\kappa$-Cl and $\kappa$-NCS, respectively. Both values are somewhat higher than those obtained in first-principles band structure calculations \cite{kand09,kore14}. It would be highly interesting to revise the calculations, taking into account our results on the SdH frequencies. At the same time, in line with the theoretical predictions, for the $\kappa$-NCS salt the $t^{\prime}/t$ ratio is significantly higher than for $\kappa$-Cl. This result clearly demonstrates the dominant role of the spin frustration in the ``chemical pressure'' effect within, at least, the present pair of $\kappa$ salts with different anions. It is interesting to perform a similar study on other $\kappa$ salts, in particular, on the metallic salt with X\,= Cu[N(CN)$_2$]Br isostructural to $\kappa$-Cl and on the spin-liquid candidate with X\,= Cu$_2$(CN)$_3$.

Finally, our analysis of the SdH frequencies  clearly reveals a considerable pressure effect on the spin frustration. For both salts, the $t^{\prime}/t$ ratio increases by $\gtrsim 20\%$ within the studied pressure range. Thus, one has to take into account the influence of pressure on both the electronic correlations and the magnetic ordering instability when studying the electronic phase diagram of our materials and, possibly, of the other $\kappa$ salts.

\begin{acknowledgments}
We are thankful to K. Kanoda, P. Reiss, and V. Zverev for stimulating and illuminating discussions. The work was supported by the German Research Foundation (Deutsche Forschungsgemeinschaft, DFG) via Grants No. KA 1652/5-1 and No. GR 1132/19-1.
\end{acknowledgments}


\begin{thebibliography}{101}
\bibitem [{com({\natexlab{a}})}]{comm_PhSh}%
  \BibitemOpen
  \href@noop {} {} \ \bibinfo {note} {When the Landau-level
  spacing becomes comparable to the interlayer transfer integral, the beats of
  SdH oscillations acquire an additional field-dependent phase shift, which
  breaks the exact $1/B$-periodicity \cite{grig02b,grig03}. In the layered
  metal $\beta$-(BEDT-TTF)$_2$IBr$_2$ displaying a similar anisotropy, the beat
  phase is shifted by about $10\%$ of the period at $B \sim 15$\,T
  \cite{grig02b}. Thus, a straightforward estimation of the Fermi surface
  warping from the positions of two beat nodes leads to an error $\lesssim
  10\%$.}\BibitemShut {Stop}%
\bibitem [{com({\natexlab{b}})}]{comm_F0}%
  \BibitemOpen
  \href@noop {} {} \bibinfo {note} {For the $\kappa$-Cl salt
  the zero-pressure value was estimated from the linear extrapolation of the
  $F_{\beta}(p)$ data in the $p$ interval 20 to 200\,MPa, see
  inset in Fig.\,\ref{F_P}(a).}\BibitemShut {Stop}%
\bibitem [{com({\natexlab{c}})}]{comm_rang}%
  \BibitemOpen
  \href@noop {} {} \bibinfo {note} {Three lowest-pressure
  points for $\kappa$-Cl, taken at $p < 40$\,MPa, were excluded from fitting. These
  points belong to the phase-coexistence region of the phase diagram, where an
  anomalous acceleration of the pressure dependence is observed \cite{ober23}.
  Consistently, these lie above the fitting curve.}\BibitemShut {Stop}%
\bibitem [{Com()}]{Comment_mass}%
  \BibitemOpen
  \href@noop {} {}\bibinfo {note} {The band cyclotron mass is ultimately
  determined by the electron density of states on the Fermi level
  \cite{meri00a}, which is quite sensitive to the details of the band-structure
  calculations and may differ by up to $30\%$ in different works, cf., e.g.,
  Refs. \cite{xu95} and \cite{ferb14}.}\BibitemShut {Stop}%
\end{thebibliography}
%
\end{document}